\documentclass[aps,prb,reprint, superscriptaddress]{revtex4-1}
\usepackage{graphicx}

\begin{document}


\title{Fluorescence and polarization spectroscopy of single silicon vacancy centers in heteroepitaxial nanodiamonds on iridium}


\author{Elke Neu}
\affiliation{Universit\"at des Saarlandes, Fachrichtung 7.2 (Experimentalphysik), 66123 Saarbr\"ucken, Germany}
\author{Martin Fischer}
\author{Stefan Gsell}
\author{Matthias Schreck}
\email[]{matthias.schreck@physik.uni-augsburg.de}
\affiliation{Universit\"at Augsburg, Lehrstuhl f\"ur Experimentalphysik IV, 86135 Augsburg, Germany}
\author{Christoph Becher}
\email[]{christoph.becher@physik.uni-saarland.de}
\affiliation{Universit\"at des Saarlandes, Fachrichtung 7.2 (Experimentalphysik), 66123 Saarbr\"ucken, Germany}

\date{\today}

\begin{abstract}
We introduce an advanced material system for the production and spectroscopy of single silicon vacancy (SiV) color centers in diamond. We use microwave plasma chemical vapor deposition to synthesize heteroepitaxial nanodiamonds of approx. \mbox{160 nm} in lateral size with a thickness of approx. \mbox{75 nm}. These oriented 'nanoislands' combine the enhanced fluorescence extraction from subwavelength sized nanodiamonds with defined crystal orientation. The investigated SiV centers display narrow zero-phonon-lines down to \mbox{0.7 nm} in the wavelength range \mbox{730-750 nm}. We investigate in detail the phonon-coupling and vibronic sidebands of single SiV centers, revealing significant inhomogeneous effects. Polarization measurements reveal polarized luminescence and preferential absorption of linearly polarized light.
\end{abstract}


\maketitle

\section{Introduction}
In recent years, color centers in diamond have attracted research interest as candidates for solid state single photon sources. For this application they offer outstanding properties including high brightness and room temperature operation. The vast majority of experiments has been performed using single nitrogen vacancy (NV) centers, demonstrating key experiments towards quantum information applications including e.g. single photon emission\cite{Kurtsiefer2000,Beveratos2001} and coherent manipulation on optical transitions.\cite{Batalov2008} Nevertheless, as a single photon source NV centers suffer from a significant drawback, namely their broad room temperature emission bandwidth of about \mbox{100 nm} caused by strong electron-phonon coupling.\cite{Kurtsiefer2000} Thus, alternative color centers with narrow \mbox{($<$10 nm)} room temperature emission namely nickel-nitrogen complexes (NE8),\cite{Gaebel2004} nickel-silicon complexes,\cite{Steinmetz2011,Aharonovich2009} chromium related centers\cite{Aharonovich2010a,Aharonovich2009b} and silicon vacancy (SiV) centers\cite{Wang2006,Neu2011,Neu2011a} were investigated. Despite promising spectral properties, first studies on single SiV centers implanted into single crystal diamond showed unfavorably low single photon emission rates of only around 1000 counts per second (cps).\cite{Wang2006} However, more recent studies using single SiV centers contained in nanodiamonds revealed superior luminescence properties featuring emission rates up to \mbox{4.8 Mcps.}\cite{Neu2011,Neu2011a} The emission is mainly concentrated in zero-phonon-lines (ZPLs) as narrow as \mbox{0.7 nm} at approx.\ \mbox{738 nm,} rendering SiV centers very promising candidates for narrowband, bright single photon sources. In Ref.\ \onlinecite{Neu2011,Neu2011a} randomly oriented nanodiamonds containing SiV centers produced \textit{in situ} during chemical vapor deposition (CVD) growth were investigated. Spatially isolated nanodiamonds (\mbox{70-140 nm} in size) allowed for the optical addressing of bright single SiV centers. Additionally, the subwavelength size of the nanodiamonds yielded an efficient out coupling of the emitted fluorescence.\cite{Beveratos2001}

In this study, we combine the advantages of spatially isolated nanodiamonds with the defined orientation of a single crystal, enabling spectroscopy of single bright SiV centers in an environment of defined orientation. The advanced material system we use consists of (001) oriented nanodiamonds grown by heteroepitaxy on nanostructured iridium/yttria-stabilized zirconia/silicon substrates.\cite{Gsell2004,Fischer2008a} Given the shape and the uniform orientation of the crystals we coin the term 'diamond nanoislands'. The samples will be discussed in detail in section \ref{sec:samples}.
The spectral properties of ensembles of SiV centers, i.e. vibronic sidebands as well as ZPL spectral width and position, have been investigated in the literature (e.g. Ref. \onlinecite{Clark1991,Rossi1997,Sittas1996,Gorokhovsky1995,Feng1993}). The variation of the local environment in which the different members of an ensemble reside is responsible for inhomogeneous line broadening. However, very little is known about these properties for single centers and their variation with the individual crystal environment (e.g. stress, proximity to other defects). With regard to the application of SiV centers as single photon sources, these inhomogeneous effects are significant as they determine the properties of individual single photon emitters, in particular the (in)distinguishability of single photons from different emitters. In the present work, we extensively study the inhomogeneous effects for single SiV centers analyzing the spectral properties of the ZPL as well as the vibronic sideband (section \ref{sec:spek}).
 Beside the spectral properties knowledge of the orientation of the radiating dipoles of color centers is of crucial importance: First, it determines the fraction of the emission collected by the optics.\cite{Plakhotnik1995} Second, knowledge of the dipole orientation is critical for the coupling of color centers to photonic  nanostructures such as nanowires\cite{Babinec2010} or photonic crystals.\cite{Englund2010} Only one experimental work has so far addressed the emission dipole orientation of the SiV center.\cite{Brown1995} We here present detailed investigations on absorption and emission dipoles of single SiV centers via polarization spectroscopy in section \ref{sec:pol}.
\section{Sample preparation and experimental setup \label{sec:samples}}
Diamond nanoislands are synthesized by microwave plasma chemical vapor deposition (MPCVD). Orientation of the grains is obtained by heteroepitaxial growth on a single crystal substrate. While different substrate materials have in principle shown the potential to serve as templates for oriented diamond growth, only iridium (Ir) fulfils the requirements for an essentially complete orientation of all the diamond grains and an unrivalled high nucleation density.\cite{Brescia2008}  In addition, all experimental observations up to now indicate its inertness in the sense that Ir does not incorporate and generate any undesired luminescent centers in diamond.

The single crystal Ir films are grown on silicon (Si) via yttria-stabilized zirconia (YSZ) buffer layers. The YSZ films are prepared by pulsed laser deposition with a KrF excimer laser (pulse duration 25 ns; pulse energy \mbox{850 mJ}) on Si(001) substrates using a cylindrical ablation target with a stoichiometry of 21.4 mol\% YO$_{1.5}$. Ablation is performed at a temperature of 750$^{\circ}$C. In order to reduce the native oxide, the first 300 pulses are performed in high vacuum. During the subsequent growth the oxygen background pressure is increased to \mbox{5$\cdot$10$^{-2}$ Pa.} For more details see Ref.\ \onlinecite{Gsell2009}.
In contrast to our standard procedure \cite{Gsell2009} which aims at flat films, we modify the Ir deposition in order to generate a nanostructured film-surface with flat (001) top facets bounded by steep side faces. This rougher topography facilitates the formation of isolated nucleation areas in the subsequent bias enhanced nucleation (BEN) procedure.

\begin{figure}
\includegraphics{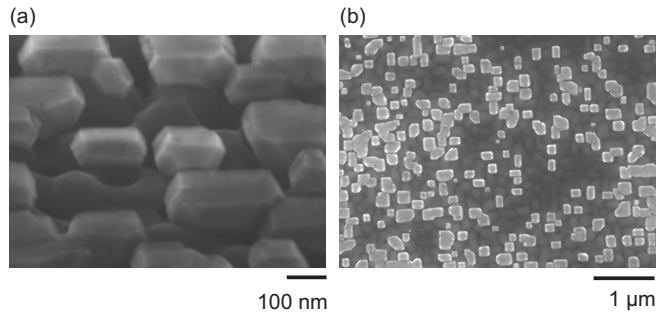}
\caption{\label{fig:rem} SEM images of diamond nanoislands (a) taken with sample tilted by 60 degrees, (b) taken at normal incidence}
\end{figure}

Diamond deposition was performed in an IPLAS reactor equipped with a CYRANNUS plasma source. During the BEN step the methane concentration in the CH$_4$/H$_2$ gas mixture is 3\%.\cite{Gsell2004} The pressure in the plasma reactor is 40 mbar, the microwave power \mbox{2000 W} and the substrate temperature 800$^{\circ}$C. A negative bias voltage of \mbox{-300 V} is applied to the substrate in order to induce an intensive ion bombardment which is necessary to generate epitaxial diamond nuclei. After switching off the bias voltage, the microwave plasma growth conditions with 0.5\% CH$_4$ in H$_2$ are maintained for \mbox{20 min.} During the growth stage the diamond nuclei transform into epitaxial diamond crystals with a mean lateral size of \mbox{160 nm} (standard deviation of \mbox{60 nm).} From SEM images (see Fig.\ \ref{fig:rem}) of the tilted sample,  the cubo-octahedral shape of the nanodiamonds with \{001\} and \{111\} faces can be clearly identified.\cite{Wild1993} We estimate a thickness of \mbox{75 nm} (standard deviation of \mbox{12 nm}). Thus, the nanoislands fulfill the requirements of subwavelength size to enable efficient fluorescence extraction with negligible total internal reflection. A low density of SiV centers is created \textit{in situ} due to plasma etching of the silicon substrates and incorporation of silicon into the growing diamond.\cite{Neu2011}

Single SiV centers are detected via confocal laser fluorescence microscopy. To excite the color centers we employ a cw tunable Titanium-Sapphire laser (Matisse, Sirah) operated at \mbox{694-696 nm.} This close to resonance excitation wavelength matches absorption bands of the SiV center.\cite{Iakoubovskii2001} The laser is focussed by a high numerical aperture microscope objective (Olympus, magnification 100x, NA 0.8). The fluorescence is collected by the same objective and separated from reflected laser light by a dichroic mirror and bandpass filters. For correlation measurements, we employ a Hanbury-Brown-Twiss setup with two avalanche photodiodes (Perkin Elmer SPCM AQRH-14) featuring a typical quantum efficiency of 70\%. Correlation measurements are performed by recording the arrival times of the photons (Pico Quant, Pico Harp, timing resolution of electronics 4 ps, photo diode timing jitter 354 ps) and calculating the correlation functions. To analyze the spectral properties of the color center fluorescence we employ a grating spectrometer (Horiba Jobin Yvon, iHr550). A grating with  600 grooves/mm here enables a resolution of approx. \mbox{0.18 nm.} All experiments were performed at room temperature.
\section{Fluorescence spectroscopy on single SiV centers in nanoislands \label{sec:spek}}
\subsection{Properties of the Zero-Phonon-Line }
\begin{figure}
\includegraphics{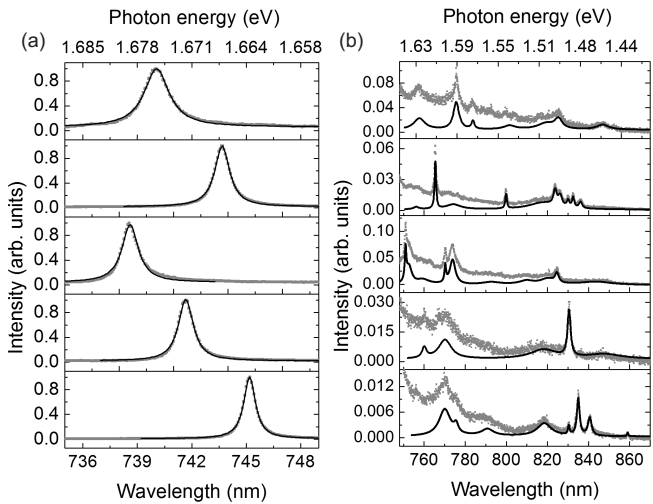}
\caption{\label{fig:spcompl} Selection of single SiV center spectra, (a) region of ZPL, black lines represent Lorentzian fits to the data, (b) sideband spectra, black lines represent the multi peak fits to the data after baseline substraction. These fits are used to determine the sideband features. All spectra are normalized to the ZPL maximum.}
\end{figure}
\begin{figure}
\includegraphics{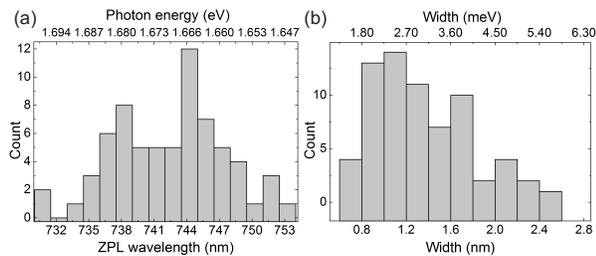}
\caption{\label{fig:zplhist} (a) Histogram of the observed ZPL positions (histogram takes into account 68 emitters) (b) Histogram of the observed ZPL width. For the histograms also spectra with multiple emitters were taken into account.}
\end{figure}

Fig.\ \ref{fig:spcompl}(a) displays 5 typical spectra of single SiV centers. Measurements of the intensity autocorrelation function (g$^{(2)}$) of the emitted fluorescence identify all color centers as bright single emitters with single photon count rates of 0.2-6 Mcps. All measured g$^{(2)}$ functions indicate three level population dynamics of the SiV centers as observed earlier.\cite{Neu2011} For the different investigated emitters we do not find a clear correlation between the g$^{(2)}$ function parameters and the emitters' spectral properties (ZPL position and width). The spectra of all investigated emitters reveal intense zero-phonon-lines (ZPLs) around \mbox{740 nm.} A histogram of the spectral positions of the ZPLs is given in Fig.\ \ref{fig:zplhist}(a). The observed positions spread over about \mbox{20 nm.} The mean value of the position is \mbox{742.6 nm} (standard deviation \mbox{5.1 nm}). The measured linewidth varies between 0.7 and \mbox{2.5 nm} with a mean value of \mbox{1.3 nm} (standard deviation \mbox{0.4 nm}). We point out that no correlation between spectral position and linewidth has been found. Thus, mechanisms broadening the ZPLs and shifting them seem to be independent. We further point out that no correlation between brightness of the emitters and line position or spectral linewidth is observed.

Evidence for environment dependent lineshifts and thus inhomogeneous broadening in ensembles is found in the literature: room temperature ZPL widths up to \mbox{15 nm} \mbox{(35 meV)} have been reported in polycrystalline diamond (PCD),\cite{Zaitsev2001} accompanied by asymmetric lineshapes and splitting of the ZPL into multiple lines in the range of \mbox{733-745 nm.} The broadening is explained by mechanical stress between crystallites.\cite{Zaitsev2001} Similarly, for SiV centers in PCD with less broadened ZPLs (\mbox{17 meV,} \mbox{7.4 nm,} \mbox{77 K}) a trend towards an asymmetric tail of the ZPL at longer wavelength is observed, hinting at a preferential red shift.\cite{Clark1995} The low temperature (10 K) linewidth observed for SiV ensembles in PCD still amounts to about \mbox{4.4 nm} (\mbox{10 meV}), indicating a significant temperature independent inhomogeneous broadening.\cite{Feng1993} The spread of the line positions of single SiV centers in our experiments is thus comparable to the inhomogeneous broadening of ensembles in PCD. The ZPL position can be shifted by stress (see Ref. \onlinecite{Sternschulte1995,Sternschulte1994}, stress in $\langle100\rangle$-direction) or by external magnetic (Zeeman-effect) or electric fields (Stark-effect). No data on the Stark-effect for SiV centers is available in the literature and the Zeeman-effect is very weak.\cite{Sternschulte1995} Thus, residual mechanical stress in the nanodiamonds is most likely the major source for the observed ZPL lineshifts.

For diamond islands on a foreign substrate several different mechanisms for stress formation have to be considered: First, stress builds up owing to the mismatch in the coefficients of thermal expansion (CTFs) between the substrate and diamond. In the present case the massive Si substrate induces a biaxial compressive stress of \mbox{-0.68 GPa} in the tiny diamond crystals when the sample is cooled down from deposition temperature to room temperature at the end of the growth process.\cite{Gsell2004} The geometric shape of isolated crystals corresponds to a truncated pyramid. In contrast to a closed diamond film, a certain fraction of the imposed stress can relax elastically. The amount of stress relaxation varies with the local position within the crystallites.\cite{Schreck2000}
Coherency stress is a second contribution specifically important in heteroepitaxial material systems. The lattice constant of Ir is 7.6\% higher than that of diamond. Pseudomorphic growth would yield an unrealistically high in-plane stress of \mbox{+89 GPa.} As a consequence misfit dislocations are introduced from the very first stage of film growth and relaxation of misfit stress occurs. In a former detailed stress analysis of a 0.6 $\mu$m diamond layer on Ir/SrTiO$_3$(001) a residual coherency stress of  \mbox{+0.9 GPa} was deduced.\cite{Schreck2000} Various measurements for 0.5-1 $\mu$m thick diamond films on Ir/YSZ/Si(001) indicate even lower values compatible with an essentially complete stress relaxation of the diamond films.\cite{Gsell2011}
Growth stress is a third source that results from the interaction of initially isolated grains when they merge during lateral growth. In PCD films nearly exclusively tensile stress is found which is attributed to attractive grain boundary forces typically explained in the framework of the grain boundary relaxation model.\cite{Windischmann1992} In Ref.\ \onlinecite{Schreck2000} a growth stress of \mbox{+0.3 GPa} was reported.

The previous considerations mainly refer to macro stress. However, as described in Ref.\ \onlinecite{Jiang1996}, during the coalescence of slightly misoriented grains, small angle grain boundaries can be substituted by disclinations and finally the initially individual crystals can continue growth as one single crystal. The coalescence process is accompanied by a strong local bending of the crystal lattice yielding high amplitudes of micro stress. Former studies\cite{Brescia2008} revealed a density of 2$\cdot10^{11}$ cm$^{-2}$ epitaxial diamond grains on Ir after BEN and 10 s growth. According to these observations we estimate that about 10 grains have merged to form the present nanoislands. Due to the extremely high activation energy for the gliding of dislocations in diamond we assume that most of the dislocations are still present in the islands of our sample. The associated micro stress fields are locally experienced by the optical centers and induce the observed distribution of ZPL positions.

For SiV centers to our knowledge there exists only one measurement of stress shift rates in the literature: in Ref.\ \onlinecite{Sternschulte1995,Sternschulte1994} stress shifting and quenching of the SiV fine structure have been investigated at low temperatures and for compressive stress applied in $\langle100\rangle$-direction. Shift coefficients in the range of -9 to 8.6 meV$\,$GPa$^{-1}$ have been determined for different fine structure components. However, there are neither measurements of lineshift coefficients at room temperature nor investigations of lineshifts for stress applied in other directions. Thus, we can state that the observed large spread in ZPL positions is due to the locally varying micro stress fields but absent knowledge of both lineshift coefficients and stress orientation preclude determination of the stress field magnitude.
We point out that the observed significant susceptibility of SiV centers to environmental changes also potentially offers controlled wide range tunability of ZPL wavelengths, already demonstrated for chromium related centers via electric fields.\cite{Muller2011} In the presence of inhomogeneous effects, resonance tuning is crucial for quantum information applications considering production of indistinguishable photons.

\subsection{Vibronic coupling and vibronic sideband features }
Even at room temperature, vibronic sideband emission of the SiV center is weak compared to the ZPL, as seen in Fig.\ \ref{fig:spcompl}(b), thus disclosing weak electron-phonon coupling. This observation is consistent with a trend reported in the literature, i.e. generally color centers involving heavier impurities tend to exhibit low electron-phonon coupling.\cite{Zaitsev2000} The strength of the electron-phonon coupling is measured by the Huang-Rhys-Factor $S$. The intensity $\vert $M$_{0n} \vert ^2 $ of the $n$ phonon sideband is given by: \cite{Davies1981}
\begin{equation}
\vert \textrm{M}_{0n} \vert ^2=S^n\frac{e^{-S}}{n!}
\end{equation}
For SiV center ensembles $S=0.24\pm0.02$ is reported in absorption  (Ref.\ \onlinecite{Collins1994}), while for single SiV centers in nanodiamonds fluorescence spectra reveal $S=0.13-0.29$.\cite{Neu2011}

\begin{figure}
\includegraphics{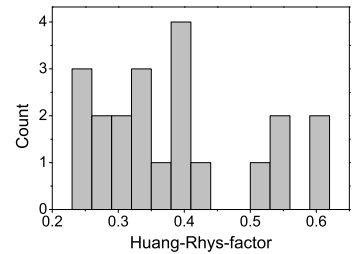}
\caption{\label{fig:hrshist} Histogram of calculated Huang-Rhys Factors taking into account emission up to \mbox{875 nm.} For the calculation baseline corrected data has been used.}
\end{figure}
Fig.\ \ref{fig:hrshist} shows a histogram of the room temperature Huang-Rhys-Factors estimated from the fluorescence spectra of single SiV centers after correcting for broadband background fluorescence originating from the nanoislands. The observed Huang-Rhys-Factors are slightly higher than reported for ensembles, the mean value amounts to 0.38 with a standard deviation of 0.12. For the majority of the SiV centers more than 70\% of the fluorescence is concentrated in the ZPL, thus rendering the SiV center especially suitable as a low bandwidth emitter. We point out that we observe 5 emitters with noticeably higher Huang-Rhys-Factor ($S>0.5$). As discussed in the following, the Huang-Rhys-Factors might be overestimated due to misinterpretation of electronic transitions as vibronic sidebands. We emphasize that we find no correlation between the ZPL position and the Huang-Rhys-Factor.\\
 As clearly visible from Fig.\ \ref{fig:spcompl}(b) the sideband structure of individual SiV centers varies significantly. In the following we analyze the origin of the sideband features as well as the variations. For the measured average Huang-Rhys-Factor of $S=0.38$ the relative intensities compared to the ZPL are 38\% for the one phonon sideband and only 7\% for the two-phonon sideband. Thus, we do not expect to observe sidebands related to two-phonon processes for SiV centers at room temperature and analyze the observed sidebands in terms of one phonon processes in the following. Electronic transitions of color centers couple to two types of vibrational modes: lattice modes corresponding to vibrations of the undisturbed diamond lattice and local or quasi-local modes. The latter are specific of the defect, representing vibrations involving the defect and its neighboring carbon atoms.\cite{LinChung1994}
Coupling to lattice modes is governed by the phonon-density of states of the diamond lattice, which has been calculated and measured in the literature (e.g. Ref.\ \onlinecite{LinChung1994,Wehner1967,Zaitsev2000,Windl1993}). Electronic transitions predominantly couple to phonons with wave-vectors at high symmetry points of the Brillouin zone (critical points).\cite{Feng1993} A list of the phonon energies corresponding to these critical points has been determined in Ref. \onlinecite{Solin1970} and is displayed in table \ref{tab:sivsidebands}. The phonon density of states has a sharp high energy cut off at \mbox{165 meV} and diminishes strongly below approx.\ \mbox{70 meV}.\cite{LinChung1994,Zaitsev2000,Wehner1967,Windl1993} Therefore, all sideband features below \mbox{70 meV} or above \mbox{165 meV} are attributed to local modes. Local modes within the energy range of lattice phonons are referred to as quasi local. The vibronic sidebands of the SiV center were experimentally examined in several publications as summarized in table \ref{tab:sivsidebands}.
 Simulations using estimated force constants for the diamond lattice and the SiV defect in Ref.\ \onlinecite{LinChung1994} indicate the existence of high energy local modes (see also table \ref{tab:sivsidebands}).

\begin{table}
\caption{\label{tab:sivsidebands}Sideband energies of the SiV center observed in the literature, all values are given in meV. The values obtained from Ref. \onlinecite{LinChung1994} are calculated values of local modes. Ref. \onlinecite{Solin1970} identifies the critical points with high phonon density (see original manuscript for identification of phonon types).}
\begin{ruledtabular}
\begin{tabular}{clllllllll}
\centering Method\tiny{$^\textrm{Ref.}$}&\multicolumn{9}{c}{Sideband energies (meV)}\\
\hline
PL 515 nm\cite{Clark1991}&&63&&&123&&154&&\\
PL 457 nm\cite{Rossi1997}&&60&&&120&&&&\\
Absorption\cite{Collins1994} &33&&&&&&&&\\
PL 488/514 nm\cite{Sittas1996}&36&64&83\footnotemark[1]&&125\footnotemark[1]&&155\footnotemark[1]&&\\
PL 488 nm\cite{Sternschulte1994}&42&65&85&&126&&153&166&\\
PL 737 nm\cite{Gorokhovsky1995}&42&64&&&125&148&155&163&\\
PL 515 nm\cite{Feng1993}&43&67&&104&129&&155&&\\
Simulation\cite{LinChung1994}&&56&80&107&&&&165\footnotemark[2]&184\footnotemark[2]\\
&&&81&&&&&168\footnotemark[2]&187\footnotemark[2]\\
critical points\cite{Solin1970}  &&70&&100&122\footnotemark[2]&138\footnotemark[2]&150\footnotemark[2]&&\\
&&&&113&133\footnotemark[2]&147\footnotemark[2]&155\footnotemark[2]&&\\
\end{tabular}
\end{ruledtabular}
\footnotetext[1]{interpreted as electronic transitions by Sittas et al.\cite{Sittas1996}}
\footnotetext[2]{several modes are predicted between these values, they are skipped for clarity.}
\end{table}
\begin{figure}
\includegraphics{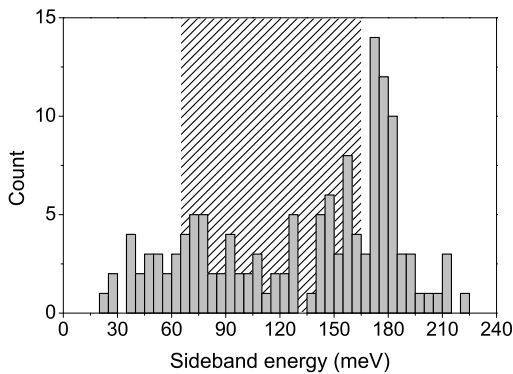}
\caption{\label{fig:sidebandshistenergy} Histogram of observed sideband features. The shaded region indicates the energy range of lattice modes. }
\end{figure}
Fig.\ \ref{fig:sidebandshistenergy} displays a histogram of the sideband energies observed for single SiV centers. The positions are derived from the fluorescence spectra by applying a multi peak Lorentzian fit to the baseline corrected spectra (see Fig.\ \ref{fig:spcompl}(b)). The observed sidebands span the whole range observed in ensembles (cf.\ Tab.\ \ref{tab:sivsidebands}), no clear concentration on distinct sideband energies is observed, except around \mbox{170-180 meV.} In accordance with the literature we observe low energy local modes around \mbox{30-40 meV,} as well as high energy local modes beyond \mbox{165 meV.} We point out that we observe high energy sidebands above \mbox{180 meV} predicted by simulations (Ref.\ \onlinecite{LinChung1994}) but not yet experimentally detected. In the energy region of lattice phonons only a weak  concentration around the critical points is observed. The relative intensity of the various sidebands changes significantly, evidencing preferential coupling to different modes (see Fig.\ \ref{fig:spcompl}(b)). We interpret the observed spread of sidebands in terms of changing electron-phonon coupling for different emitters, also indicated by the varying Huang-Rhys-Factors.

In the literature several reports indicate that sidebands and thus electron-phonon coupling strongly depend on the local environment. Sternschulte et al.\ \cite{Sternschulte1994} report that a sideband feature shifted by \mbox{166 meV} was only observed on some positions on a CVD grown homoepitaxial sample. Iakoubovskii et al.\ \cite{Iakoubovskii2001} find a correlation between the width of the Raman-peak of the CVD film, i.e. the stress present in the film and the sidebands of the SiV center: in highly stressed films broadband emission dominates over distinct sideband features.  Gorokhovsky et al.\ \cite{Gorokhovsky1995} also report that in PCD no well resolved sideband features were observed under non-resonant \mbox{514 nm} laser excitation, indicating averaging over centers with differing sideband spectra. In contrast, for resonant excitation of the same sample with \mbox{737 nm} a defined sideband structure evolves. The authors in Ref.\ \onlinecite{Gorokhovsky1995} attribute this to the excitation of a sub-ensemble with ZPL at the excitation laser wavelength. This sub-ensemble within the inhomogeneously broadened line exhibits defined and identical sideband features, while the sideband structure is averaged out upon excitation of all inhomogeneously broadened SiV centers. If the resonant excitation is tuned to longer wavelengths, the most prominent sideband feature at 64 meV shifts to lower phonon energies (\mbox{2.5 meV,} excitation tuned from \mbox{736-739 nm).} The shifting behavior for a higher energy sideband at \mbox{125 meV}, however, shows no clear trend. Thus, distinct sidebands may respond differently to environmental changes that shift the ZPL accompanied also by changes in the linewidth of the phonon replicas.\cite{Gorokhovsky1995}  We here observe the varying sideband spectra directly via observation of single emitters. Thus, we exclude inhomogeneous effects in sub ensembles. The observed variety of single emitter sideband spectra is in accordance with previous ensemble observations.

\begin{figure}
\includegraphics{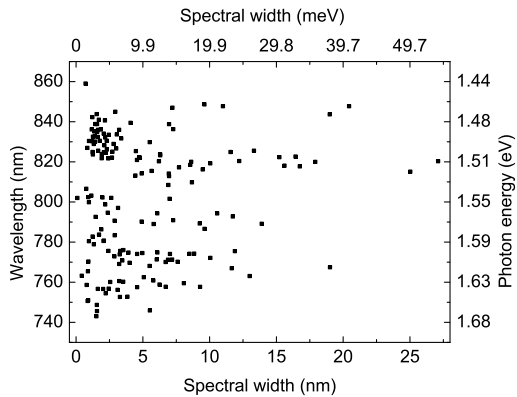}
\caption{\label{fig:sidebandwidthwavel} Width of the observed sideband features plotted versus wavelength (photon energy) of the features. Note that the scale for the width in meV gives only approximate values calculated for an absolute energy of 1.57 eV  }
\end{figure}

As visible from Fig.\ \ref{fig:spcompl}(b) some SiV centers show remarkably narrow emission lines in the sideband region. Specifically, around 825-845 nm individual SiV centers even show multiple narrow lines. Fig.\ \ref{fig:sidebandwidthwavel} displays the observed features plotted versus their width, indicating an accumulation of narrow features in the spectral region from 825-845 nm. The narrow linewidth suggests that these features are due to electronic transitions rather than vibronic sidebands. However, the identification of sidebands and electronic transitions is not always clear as for defects involving heavy impurities sideband features due to local modes might be as narrow as the ZPL itself.\cite{Zaitsev2000} Nevertheless, in low temperature experiments Sittas et al.\ \cite{Sittas1996}  observe a significant spectral narrowing of the three highest energy sidebands of the SiV \{776 nm (83 meV), 797 nm (125 meV) and 812 nm (155 meV)\} and thus attribute them to purely electronic rather than vibronic transitions. As in previous studies of SiV centers photon correlation measurements provided evidence for three level population dynamics,\cite{Wang2006,Neu2011,Neu2011a} one might expect a second purely electronic transition. In addition, very recent studies report absorption of SiV ensembles between 830-850 nm.\cite{Dhaenens2011}  Therefore, we tentatively attribute the observed features to electronic transitions from a second excited state to the ground state. We point out that spectral features above 900 nm are strongly suppressed in our setup for technical reasons (dichroic transmission, spectrometer efficiency): The detection efficiency is only approx. 0.25\% of the efficiency at 750 nm. Thus, we would not expect to observe e.g. the 946 nm line recently reported as a further electronic transition of the SiV center.\cite{Dhaenens2011}

\section{Polarization spectroscopy \label{sec:pol}}
\begin{figure}
\includegraphics{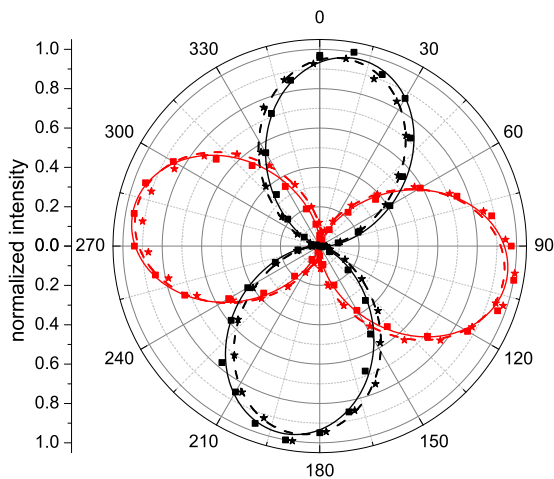}
\caption{\label{fig:zusammenfpol} (Color online) Polarization dependent absorption measurement (data points: filled squares, fit: solid lines) and polarization of emission (data points: filled stars, fit: dashed line) of two single SiV centers (red curves: center 1, black curves: center 2)}
\end{figure}
In this section, we investigate the polarization dependent absorption of the SiV centers, as well as the polarization properties of the emitted fluorescence light. For the first measurement a half-wave plate is used to rotate the excitation laser polarization while recording the emitted fluorescence intensity. To ensure that the fluorescence is proportional to the absorbed intensity, excitation powers far below saturation are employed. To determine the polarization of the emitted light a linear polarization analyzer is rotated in the detection path.
Different absorption processes are possible for color center excitation. For defect to band excitation electrons excited to the conduction band relax via defect states. As the ground state of the SiV center was reported about \mbox{2.05 eV} (Ref. \onlinecite{Iakoubovskii2000b}) below the conduction band edge, we exclude this path for our \mbox{1.78 eV} (695 nm) excitation. The absorbing transition can also be provided via transitions to higher vibronic states in the excited state or higher excited electronic states of the color center, thus identification of the absorbing transition is unclear. If the SiV center behaves like a single dipole in absorption one expects a sinusoidal characteristics with minima close to zero for excitation with linearly polarized light, as no light is absorbed if its polarization is perpendicular to the dipole axis. The visibility $V$ \begin{equation}
      V=\frac{I_{max}-I_{min}}{I_{max}+I_{min}}\,.
\end{equation} of a single dipole in absorption amounts to 100\%. Additionally, the absorption only addresses the dipole component in the sample plane (i.e. plane perpendicular to the excitation laser propagation direction, here (001) plane of the diamond).\cite{Ha1999} Dipoles oriented in that plane show maximum absorption, while dipoles oriented perpendicularly are effectively not excited. All SiV centers investigated show very high visibility $V$ in absorption between 90\% and 100\%. Thus, we conclude that SiV centers exhibit a single dipole in absorption ($\pi$ type transition). The deviations from 100\% visibility are attributed to experimental imperfections. First, to determine the contrast in the polarization dependent excitation curves we correct for background luminescence of the diamond material. As the background exhibits spatial variations in the vicinity of the defects, this procedure introduces an uncertainty of 5-10\%. Second, the dichroic mirror used to separate excitation laser light and color center luminescence induces polarization changes, leading to an orientation dependent loss of linear polarization: for s and p polarization reflected laser light maintains 100\% polarization visibility, for 45 degrees we measure a reduction to about 90\%.

\begin{figure}
\includegraphics{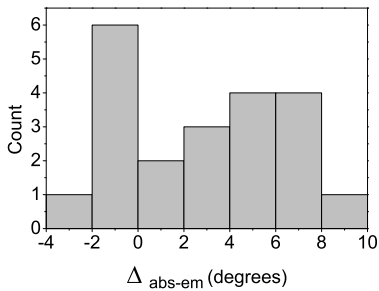}
\caption{\label{fig:absemdev} Deviation of measured emission and absorption dipole orientation (in sample plane) as derived from polarization dependent measurements. The mean value of the deviation is 2.3 degrees. The maximum deviation is 8.2 degrees.}
\end{figure}
Similar to the absorption, we observe high polarization visibility in the range of 86\% to 100\% for the fluorescence, thus evidencing linearly polarized emission. The absence of depolarization hints at vanishing influences of reorientation processes in the excited state that would lead to depolarization of the emitted light.\cite{Clark1962}
Furthermore, from the measured polarization dependence we infer that emission and absorption dipoles of the SiV centers are parallel within experimental error. Fig.\ \ref{fig:absemdev} gives a histogram of the observed deviations, the maximum deviation is 8.2 degrees, the mean value is 2.3 degrees. For a purely statistical measurement error we would expect the mean value of the deviations to be zero. We thus interpret this result as an offset of 2 degrees of the polarization scale for absorption and emission. We point out that a parallel orientation of absorption and emission dipoles is very common e.g. for organic molecules (Ref. \onlinecite{Ha1999}) and has also been considered for other vacancy based color centers for $\pi$-type absorption and emission dipoles.\cite{Bergman2007}  Recent studies on chromium related centers also found parallel orientations for emission and absorption dipoles, however, with significantly varying orientations in the diamond lattice.\cite{Muller2011}
\begin{figure}
\includegraphics{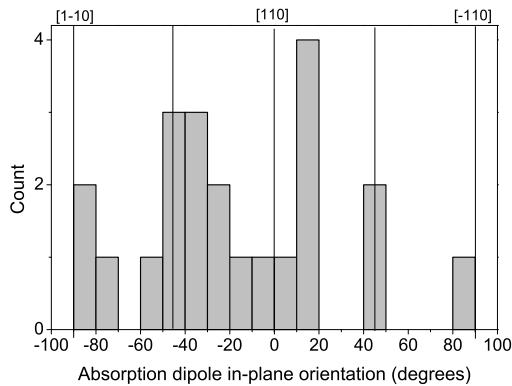}
\caption{\label{fig:absdipoleorient} Histogram of measured SiV center absorptions dipole orientations in the sample plane (001) }
\end{figure}
In the following we aim at interpreting our polarization measurements in terms of orientation of the color center dipole with respect to crystal axes and electronic states involved in the transition. For this interpretation we have to consider that the absorption measurements only reveal the dipole component in the sample plane. On the other hand, care has to be taken in interpreting emission polarization: First, the presence of the metal surface might distort the polarization of the emitted light. Second, due to imaging through a high NA objective, loss of polarization contrast occurs for linearly polarized emission.\cite{Fourkas2001} For a dipole in the sample plane we would expect 100\% visibility, whereas for a dipole perpendicular to that plane polarization contrast fully vanishes, a 45 degrees tilt yields 70\% visibility (NA 0.8). Thus, in principle, assuming a single dipole, one may use the visibility as well as the position of maxima and minima to determine the three-dimensional orientation of the dipole.\cite{Fourkas2001} Nevertheless, the considerations in Ref.\ \onlinecite{Fourkas2001} only hold for a dipole in an isotropic medium. SiV centers in nanodiamonds on an Ir surface constitute a highly anisotropic situation, leading to a strongly modified radiation pattern close to the metal surface.\cite{Neu2011} Taking also into account the issue of background substraction, we therefore consider the polarization contrast non reliable to determine the tilt of the dipole with respect to the sample plane in our case. Fig.\ \ref{fig:absdipoleorient} displays a histogram of the orientations of the SiV dipole in the sample plane deduced from absorption measurements. We note that the nanoislands $\langle110\rangle$ crystal directions in the (001) plane align with the sample edges and we identify a 0 degrees orientation of the azimuthal angle with the [110] direction.

Generally, point defects in diamond may have their highest symmetry axis oriented along $\langle100\rangle$, $\langle110\rangle$ or $\langle111\rangle$ axes.\cite{Kaplyanskii1963} For $\langle100\rangle$ oriented defects 3 equivalent sites have to be considered, while there are 6 equivalent sites for $\langle110\rangle$ and 4 equivalent sites for $\langle111\rangle$ oriented defects.\cite{Brown1995}  Due to symmetry considerations transition dipoles are either parallel to the high symmetry axis (z-dipole) or perpendicular (x,y-dipoles).\cite{Brown1995} Based on measurements of polarized luminescence Brown et al.\ \cite{Brown1995} identify a z-dipole for the ZPL transition of the SiV center. Following their line of argument, we assume a z-dipole for the interpretation of the orientation data. For our (001) oriented sample, one expects dipole orientations in the sample plane of -45 and 45 degrees for $\langle100\rangle$ oriented transition dipoles, while for $\langle111\rangle$ oriented defects orientations -90 degrees, 0 and 90 degrees occur. $\langle110\rangle$ oriented defects on the other hand, lead to measured dipole orientations of -90, -45, 0, 45 and 90 degrees. Whereas we observe a certain scatter in orientations, the measured data, nevertheless, best match a $\langle110\rangle$ defect orientation and exclude $\langle100\rangle$ and $\langle111\rangle$ orientations (see Fig.\ \ref{fig:absdipoleorient}). SEM images imply that the spread of dipole orientations is not due to misorientation (twist and tilt) of the diamond nanoislands. In summary, our findings are similar to those of Brown et al.\ \cite{Brown1995}, identifying the SiV center as a $\langle110\rangle$ oriented defect with a z-dipole. Brown et al. assign the defect to be of monoclinic I or rhombic I symmetry.\cite{Brown1995}
We now compare our observations to theoretical models discussed in the literature. Goss et al.\cite{Goss1996,Goss2007} model the SiV center as a negatively charged defect with D$_{3d}$ symmetry, the silicon atom being located in a so called split-vacancy configuration. Their calculations identify a ${}^2E_g \rightarrow {}^2E_u $ transition as the 1.68 eV ZPL. At low temperature the ZPL splits into 4 components.\cite{Clark1995, Sternschulte1994} The splitting is attributed to a Jahn-Teller-effect in this model. The light emitted by a ${}^2E_g \rightarrow {}^2E_u $ transition  (D$_{3d}$), however, is unpolarized according to Kaplyanskii.\cite{Kaplyanskii1963} Thus, this model does not fit our experimental observation. An alternative model introduced by Moliver in Ref.\ \onlinecite{Moliver2003} on the other hand assumes the center being in the neutral charge state (SiV$^0$). The silicon atom is shifted off center along the $[111]$ direction yielding $C_{3\nu}$ symmetry. The splitting is explained in terms of a tunneling of the Si atom between equivalent sites. The 1.68 eV ZPL is associated with a ${}^3A_2^\star,{}^3E^\star\rightarrow {}^3A_2$ transition. Here the transition between $A_2$ states would be linearly polarized.\cite{Kaplyanskii1963} Thus, our polarization data would support this model of the SiV$^0$. Nevertheless, the orientation data does not fit the symmetry proposed therein. Furthermore, the interpretation as SiV$^0$ defect is questionable, as the SiV$^0$ defect has recently been identified as the source of 1.31 eV emission using optical and electron paramagnetic resonance techniques.\cite{Dhaenens2011} In summary, the measured dipole orientations imply a lower symmetry of the defect than predicted by prevalent theoretical models and assignment of transition polarizations and charge states seem questionable. To provide a reliable matching of theoretical prediction and experimental data a more precise determination of the dipole orientation as well as further theoretical work are necessary.
\section{Conclusion}
We extensively investigated the spectral properties of single SiV centers produced \textit{in situ} during CVD growth. As advanced material system we employ \mbox{160 nm} large, \mbox{75 nm} thick heteroepitaxially grown nanoislands on Ir/YSZ/Si substrates, thus combining enhanced fluorescence extraction and defined crystal orientation. We find intense, narrow \mbox{(0.7-2.5 nm)} ZPLs at around \mbox{740 nm.} Residual stress in the nanoislands leads to a spread in peak position of about \mbox{20 nm.}  The SiV centers display low electron-phonon coupling with a mean Huang-Rhys-Factor of 0.38. Individual sideband spectra vary significantly in accordance with previous ensemble observations. Between 825-845 nm an accumulation of narrow peaks is observed, tentatively attributed to a second electronic transition. Polarization measurements show preferential absorption of linearly polarized light, linearly polarized emission as well as evidence for orientation of the center's dipole along the $\langle110\rangle$ direction. These findings are in accordance with previous experimental results but contradictory to current theoretical models for the SiV center. The spectroscopic properties deduced here support the suitability of SiV centers as narrow bandwidth, high brightness room temperature single photon emitters.

\begin{acknowledgments}
We acknowledge funding by the DFG  and the BMBF (EPHQUAM 01BL0903).
SEM measurements were performed by J. Schmauch (UdS).
\end{acknowledgments}
%

\end{document}